\documentclass[aps,pra,reprint,10pt,a4paper]{revtex4-1}
\usepackage[utf8]{inputenc}
\usepackage[sc,osf]{mathpazo}
\usepackage{amsmath}
\usepackage[T1]{fontenc}
\usepackage{latexsym}
\usepackage{amssymb}
\usepackage[colorlinks=true,citecolor=blue,urlcolor=blue]{hyperref}
\usepackage{color}
\usepackage{graphics,epstopdf}
\usepackage{soul}
\usepackage[demo]{graphicx}
\usepackage{capt-of}
\usepackage{lipsum}
\usepackage{adjustbox}
\usepackage[normalem]{ulem}
\usepackage[table,xcdraw]{xcolor}
\usepackage{braket}
\usepackage{physics}
\usepackage{ragged2e}
\usepackage{comment}
\usepackage[font=small,labelfont=bf,justification=justified,singlelinecheck=false]{caption}

\begin{document}

\title{
Production of genuine multimode entanglement  in circular waveguides with long-range coupling}

\author{Tonipe Anuradha$^{1}$, Ayan Patra$^{1}$, Rivu Gupta$^{1}$, Amit Rai$^{2}$, Aditi Sen(De)$^{1}$}

\affiliation{$^{1}$ Harish-Chandra Research Institute,  A CI of Homi Bhabha National Institute, Chhatnag Road, Jhunsi, Prayagraj - 211019, India }
\affiliation{$^{2}$ School of Physical Sciences, Jawaharlal Nehru University, New Delhi 110067, India}

\begin{abstract}
 
Starting with a product initial state, squeezed (squeezed coherent) state in one of the modes, and vacuum in the rest, we report that a circular waveguide comprising modes coupled with varying coupling strength is capable of producing genuine multimode entanglement (GME), quantified via the generalized geometric measure (GGM). We demonstrate that for a fixed coupling and squeezing strength, the GME content of the resulting state increases as the range of couplings between the waveguides increases, although the GGM collapses and revives with the variation of coupling strength and time.
The advantage of long-range coupling can be emphasized by measuring the area under the GGM curve, which clearly illustrates growing trends of GME with the increasing range of couplings. Moreover, long-range couplings help in generating a higher GGM for a fixed coupling strength.
We analytically determine the exact expression of GGM for systems involving an arbitrary number of modes, when all the modes interact with each other equally. The entire analysis is performed in the phase-space formalism. We manifest the constructive effect of disorder in the coupling parameter, which promises a steady production of GME, independent of the coupling strength.

\end{abstract}
\maketitle

\section{Introduction}
\label{sec:intro} 

Continuous variable systems, characterized by position and momentum quadratures \cite{Serafini_2017}, are one of the potential platforms for the experimental realization of a wide range of quantum information processing tasks. Notable ones include quantum communication protocols \cite{furusawa1998, bowen2003, mizuno2005} with or without security \cite{li2002, lance2005}, quantum cloning machine \cite{andersen2005}, and the preparation of cluster states \cite{Yoshikawa_APL_2016} essential for building one-way quantum computer \cite{Raussendorf_PRL_2001}. One of the key resources required to design these quantum protocols is multimode entanglement \cite{Horodecki_RMP_2009}. Therefore, the generation of entanglement in physical substrates \cite{Masada_NP_2015, Lenzini_SA_2018, Larsen_NPJ_2019},  its detection \cite{Simon_PRL_2000, Duan_PRL_2000, Giedke_PRA_2001, vanLoock_PRA_2003, Armstrong_NP_2015, Qin_NPJ_2019}, and quantification \cite{Adesso_PRA_2004, Braunstein_RMP_2005, Guhne_PR_2009} have attracted lots of attention.

Coupled optical waveguides in a one-dimensional array turn out to be an efficient method to manipulate light \cite{Takesue_Optica_2008, Camacho_Optica_2012, Das_PRL_2017, Kannan_SA_2020, Zhang_Nature_2021} or to simulate quantum spin models via optics \cite{Hunh_PNAS_2016, Bello_PRX_2022}.
Thus they have emerged as suitable candidates for performing continuous time random walks \cite{perets_2008,peruzzo_2010_quantum}, Bloch oscillation \cite{Morandotti_PRL_1999, Pertsch_PRL_1999, Sapienza_PRL_2003, Dreisow_OL_2011}, Anderson localization \cite{Martin_OE_2011}, quantum computation \cite{Fu_SPIE_2003, Politi_S_2008, Paulisch_IOP_2016}, optical simulation \cite{keil_2015_optical} and generation of entangled states \cite{rai_2010_quantum}. 
Various studies have utilized different linear waveguide array models to detect continuous variable entanglement  via the van Loock and Furusawa inequalities \cite{vanLoock_PRA_2003, Rai_PRA_2012}, and to quantify entanglement between two modes using logarithmic negativity \cite{Barral_PRA_2017, Barral_PRA_2018, Asjad_PRAl_2021}. More recently, the transfer of quantum states of light between modes in circular waveguide arrays has also been explored \cite{rai_2022_transfer}. 

A majority of these works are based on Hamiltonians involving couplings only between neighboring modes, popularly known as nearest-neighbor (NN) couplings,
although non-nearest-neighbor coupling is essential in some situations. For instance, in quantum information processing and quantum computation with optical waveguides, it is necessary to fabricate compact waveguide circuits to reduce the footprints of such circuits \cite{meany_2015}. When the separation between the waveguides in such circuits would keep on decreasing, or when 
the waveguide is long, the higher-order couplings must be taken into account.
Benefits of non-nearest-neighbor couplings have been shown in molecular excitation transfer \cite{Gaididei_PRE_1997}, the study of Bloch oscillations in photonic waveguide lattices \cite{Morandotti_PRL_1999, Pertsch_PRL_1999, Sapienza_PRL_2003, Dreisow_OL_2011}, the dynamics of bio-molecules \cite{Mingaleev_JBP_1999} and polymer chains \cite{Hennig_EPJB_2001}. Moreover, long-range (LR) couplings play a vital role in localization \cite{Lopez_NP_2008}, simulations \cite{Aspuru_Nature_2012} and quantum walks in waveguide systems.
More importantly, such LR couplings can be simulated and manipulated in laboratories with several physical systems including photonic waveguides \cite{Davis_OL_1996, Kevrekidis_PD_2003, Iyer_OE_2007, Szameit_OL_2009, Longhi_LPR_2009, Garanovich_PR_2012, Golshani_PRA_2013} (c.f. \cite{Jones_JOSM_1965, Estes_PR_1968}), trapped ions \cite{Porras_PRL_2004, Islam_Nature_2011} etc.

Here, we provide a technique that uses circular waveguide arrays, with long-range couplings, to produce genuine multimode entangled (GME) states from product ones. 
We point out that our work is significant since most of the earlier research works relating to continuous variable (CV) multimode entanglement involve the use of bulk optical elements, which are large and inherently sensitive to decoherence resulting in a reduction of entanglement content.
In this article,  we focus on integrated photonic waveguides which can be fabricated using femtosecond laser techniques\cite{pertsch_2004_discrete,itoh_2006_ultrafast,szameit_2010, meany_2015} and nanofabrication methods \cite{Rafizadeh_CLE_1997, Belarouci_JL_2001}, having minimal decoherence \cite{perets_2008, Dreeben_QST_2018}. These platforms guarantee a very low loss factor and are interferometrically stable, scalable, and less susceptible to decoherence, thereby ensuring robustness against noise.

In continuous variable systems, a majority of the previous works analyzed whether genuine multimode entanglement creation was successful or not \cite{Asjad_PRAl_2021, Barral_PRA_2018}, through the application of the van-Loock Furusawa inequalities \cite{vanLoock_PRA_2003}, for systems with up to five modes \cite{Rai_PRA_2012}, even though multimode entangled states are crucial for several quantum information protocols \cite{Hillery_PRA_1999, Cleve_PRL_1999, Gottesman_PRA_2000, Bruss_PRL_2004, Ishizaka_PRL_2008, Bennett_TAQC_2014}. \textcolor{black}{Going beyond detecting entanglement, 
we \textit{quantify} genuine multimode entanglement by computing the generalized geometric measure (GGM) \cite{Shimony_ANYAS_1995, Barnum_JPA_2001, Wei_PRA_2003, SenDe_PRA_2010, Das_PRA_2016, Buchholz_AP_2016} for CV Gaussian system by using phase-space formalism \cite{Roy_PRA_2020}, and explicitly show how long-range interactions are beneficial for generating genuine multimode entanglement.}
In particular, the multimode entangled state is generated using waveguides organized circularly and coupled with varying coupling strengths, where a squeezed state of light is given as input in one mode and vacuum impinges on the other modes. 
Note that it does not require nonlinear processes which are relatively more difficult to work with. We first observe that irrespective of the range of couplings, GGM collapses and revives with the variation of the coupling constant and time. 
By exploiting the symmetry of the system, we analytically arrive at the compact form of GGM when the dynamics are driven by the LR couplings having equal strengths. We illustrate that the time-varying GME content can be higher for the LR model than that of the NN model, for a fixed coupling and squeezing strength, although \emph{the maximum  GGM produced with NN coupling coincides with the one generated by waveguides having LR couplings}. \textcolor{black}{Note that the enhancement of entanglement with LR interactions as compared to the NN ones aligns with the expected outcome based on the area law of entanglement entropy although the trends and maximal GGM cannot be explained via entanglement area law. Most of the studies on entanglement area law of modes having both short and long-range interactions are for finite-dimensional systems which is not the case in the current study (see \cite{Vitagliano_NJP_2010, Movassagh_PNAS_2016} for violations of the area-law). Moreover, we provide a method to compute the GME in such systems, which is not restricted with regard to the number of involved modes and is scalable to an arbitrarily high number of modes when all-to-all interactions are implemented. 
}  \\
 

We also show that if disorder is introduced in the couplings, the oscillations in the generated quenched averaged GGM decrease at the expense of the maximum GGM content. It indicates that the generation of a non-oscillating genuine multimode entanglement can only be accomplished when there are some imperfections in the coupling strength that naturally arise during the implementation of the waveguide system. Additionally,  the quenched average GGM increases with the increase of the range of couplings involved in the evolution process. \textcolor{black}{Based on the period of oscillation of GGM with respect to the coupling strength, it is possible to obtain a bound on the disorder strength which can lead to oscillation-free GGM. }

Our paper has the following structure: Sec. \ref{sec:pre} provides a brief overview of the theoretical model for a circular array of linear waveguides, including the Hamiltonian and the input state. In Sec. \ref{sec:LR}, we explain the benefits of taking long-range couplings for creating genuine multimode entanglement in four, five, and six modes. \textcolor{black}{We analyze the scaling of the block entropy of entanglement in \ref{subsubsec:block_entropy}, which supports the computation of the GGM in Secs. \ref{subsubsec:GGM_expresion} and \ref{subsubsec:GGM-N} using long-range interactions in systems comprising an arbitrary number of modes. }
Sec. \ref{sec:disorder} explores the impact of disorder present in the coupling strength on multimode entanglement. Finally, we conclude in Sec. \ref{sec:conclu}.

\section{Design of the Waveguide setup}
\label{sec:pre}
 \begin{figure}
    \centering
    \includegraphics[width=\linewidth]{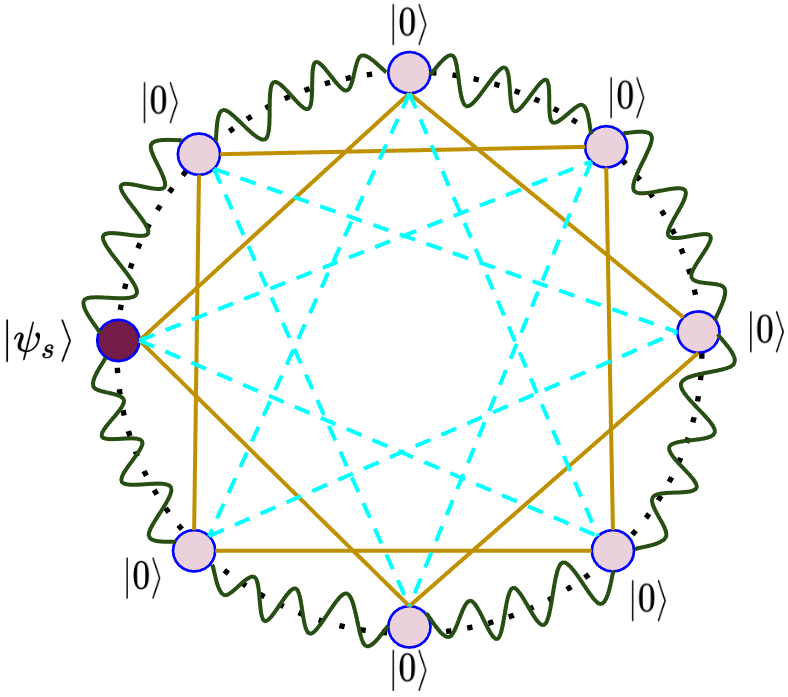}
\captionsetup{justification=Justified,singlelinecheck=false}
    \caption{Circular waveguide setup for the generation of genuine multimode entanglement between eight optical modes. The dark circle represents the mode in which the squeezed state $|\psi_s\rangle$ is given as input, whereas the light circles denote the vacuum $|0\rangle$ modes. The dark green curved lines correspond to the nearest-neighbor (NN) coupling. Long-range couplings are shown as follows: next-nearest-neighbor (NNN) with light yellow straight lines and next-to-next-nearest-neighbor (NNNN) coupling as very light dashed blue lines. For a waveguide with a large number of modes, higher levels of long-range coupling have to be incorporated. The coupling strengths of the NN and long-range couplings, in general, can be different.}
    \label{fig:schematic}
\end{figure}

Let us first introduce the model which describes the evolution of the product input state to a genuinely multimode entangled state. The system comprises $N$ identical waveguides arranged in a circular configuration and coupled to each other, with varying coupling strength (for a schematic description of the system, see Fig. \ref{fig:schematic} for $N = 8$).  The Hamiltonian that governs the couplings of the $N$ modes within the system is represented by
\begin{equation}
\begin{split}
    \hat{H}=&\sum_{i=1}^{[\frac{N}{2}]-1}\hbar J'_i\sum_{j=1}^{N}( \hat{a}^\dagger_j\hat{a}_{j+i} +H.c.)+\\&\frac{1}{1+\frac{1}{2}(1+(-1)^N)}\hbar J'_{[\frac{N}{2}]}\sum_{j=1}^{N}( \hat{a}^\dagger_j  \hat{a}_{j+[\frac{N}{2}]} + \text{H.c.}),
   \label{eq:H_n}
\end{split}
\end{equation}
where an increasing $i$ indicates an increasing range of couplings. Here $N+j\equiv j(mod~N)$, $\hat{a}_j$ and $\hat{a}_j^\dagger$ are the respective bosonic annihilation and creation operators for the $j$-th mode, H.c. stands for the Hermitian conjugate,  $J'_i$ denotes the coupling strength or coupling constants between waveguide modes with $J'_1 = J'$ and $J'_i = n_i J'$ for $i \geq 2$ and we consider $\hbar=1$. Thus $J'$ represents the strength of the nearest-neighbor (NN) coupling. The long-range coupling is introduced by making $n_i >0$ for $i \geq 2$. We must note that  $n_i \neq 0$, if and only if $n_j \neq 0$ $\forall j < i$, with the condition $0< n_i \leq 2$ \cite{Dreisow_OL_2008}. The second term in Eq. \eqref{eq:H_n} takes care of the longest range of couplings between modes. 

\textcolor{black}{The Hamiltonian under consideration leads to the generation of genuine multimode entanglement in any experimental setup capable of simulating the proposed evolution, even with the help of bulk optical elements. In our study, we utilize evanescently coupled waveguides that are fabricated using the femtosecond laser direct-writing method, as demonstrated in Ref. \cite{szameit_2010}. It should be noted, however, that with an increase in the number of modes, implementation of the protocol through bulk optical elements would increase the scope of decoherence. On the other hand, waveguide setups provide decoherence resistance and interferometric stability \cite{meany_2015, perets_2008} even for arbitrary sizes and thus also provide scalability. Furthermore, higher-order couplings have already been engineered in such setups \cite{Dreisow_OL_2008}, thereby making them suitable candidates for the efficient realization of the proposed scheme involving long-range interactions for entanglement generation.}

\textbf{Note $\mathbf{1}$.} 
The time evolution operator corresponding to the Hamiltonian in Eq. \eqref{eq:H_n} is given by $\exp(-i \hat{H}t)$. Therefore, upon evolution, the final state of the waveguide system contains terms of the form $J' t$. We relabel such parameters as $J$, representing  
the coupling strength or the coupling parameter and the range of the couplings is tuned with \(n_i\). Thus, the variation with respect to $J$ also represents the variation in time. Moreover, note that $t=z\mu/c$, where $\mu$ is the refractive index for the waveguide mode, which relates the time duration $t$ to the propagation distance $z$.

In order to create a genuine multimode entangled state from a fully product state, we study the dynamics induced by the aforementioned couplings to identify the optimal configuration of the waveguide system.
In particular, one of the modes, say, the first mode, is chosen to be a single-mode squeezed state, $\ket{\psi_{\text{s}}}=\exp(\frac{1}{2}(\xi^* \hat{a}_j^2-\xi \hat{a}_j^{\dagger^2}))\ket{0}$ with $j$ being the input site, the squeezing parameter is $\xi= se^{i\theta}$, where $s$ is the squeezing strength and $\theta$ represents the squeezing angle. The rest of the modes are in the vacuum state, $|0\rangle$, i.e., the $N$-mode initial state takes the form as
\begin{equation}
    |\psi\rangle_{\text{in}} = \ket{\psi_{\text{s}}} \otimes |0\rangle^{\otimes N - 1}.
    \label{eq:initial_state}
\end{equation}
The covariance matrix corresponding to the above initial state has the form,
\begin{eqnarray}
   \nonumber \Xi_i=\!\!\!\!\!&&\frac{1}{2}\Bigg[\begin{pmatrix} 
        \cosh 2s + \cos \theta \sinh 2s & \sin  \theta \sinh 2s \\
        \sin \theta \sinh 2s & \cosh 2s - \cos \theta \sinh 2s
    \end{pmatrix} \\ 
   &&\oplus \mathbb{I}^{\oplus N - 1}\Bigg],
    \label{eq:initial_cov-mat}
\end{eqnarray}
where $\mathbb{I} = \text{diag}(1,1)$ is the $2 \times 2$ identity matrix. Note that due to the periodicity present in the model, the position of the mode in which the input squeezed state is taken cannot alter the multimode entanglement content of the final state.

The symplectic formalism is used to analyze the evolution of the Gaussian input state and to characterize its entanglement (see Appendix \ref{app:CV} for details of the analytical formalism). The covariance matrix corresponding to the initial state of the system is denoted as $\Xi_{\text{in}}$. The  final state of the system, upon evolution, is characterized by ${\Xi_f}=S_H {\Xi_{in}} S_H^T$, where $S_H$ is the symplectic transformation of the waveguide Hamiltonian, as defined in Appendix \ref{app:CV}, for which the generalized geometric measure is computed (see Appendix. \ref{app:CV_GGM} for the computation of GGM for a pure CV Gaussian state). In Appendix \ref{sec:app_3-mode}, we present the simplest model involving a state with three modes propagating through circularly coupled waveguide modes that have only the nearest-neighbor coupling. It is important to emphasize here that such treatment provides the possibility to address this problem involving an arbitrary number of modes.

\textbf{Remark $\mathbf{1}$}. Instead of the squeezed state, if one considers a coherent state as the input,  such a generation of multimode entanglement is not possible. This can be explained by considering the covariance matrix of the coherent state, which is nothing but $\frac{1}{2} \mathbb{I}$. Thus, in this scenario, the input covariance matrix reduces to $\Xi_{i} = \frac{1}{2} \mathbb{I}^{\oplus N}$ and the final state of the system is denoted by a covariance matrix proportional to the identity matrix.  
Thus, starting from a product state, we again end up with a product state after evolution and the entanglement generation cannot occur. 

\textbf{Remark $\mathbf{2}$}. With a squeezed coherent state as input in one of the modes (and vacuum in the rest) of Eq. \eqref{eq:initial_state}, the entanglement generated among the $N$-modes is the same as that obtained via an input squeezed state.

\section{Advantage of long-range coupling in Entanglement creation}
\label{sec:LR}

In typical waveguide systems studied in the literature, only the NN couplings are considered, while higher-order couplings lead to bosonic Hamiltonians with LR couplings as in Eq. \eqref{eq:H_n} which will be the main focus of this work. The motivation behind such consideration is the fact that in several physical systems, especially quantum spin models, LR couplings have been shown to typically create highly multimode entangled states, which serve as resources for quantum information processing tasks.
 \begin{figure*}
    \centering
    \includegraphics[width=\linewidth]{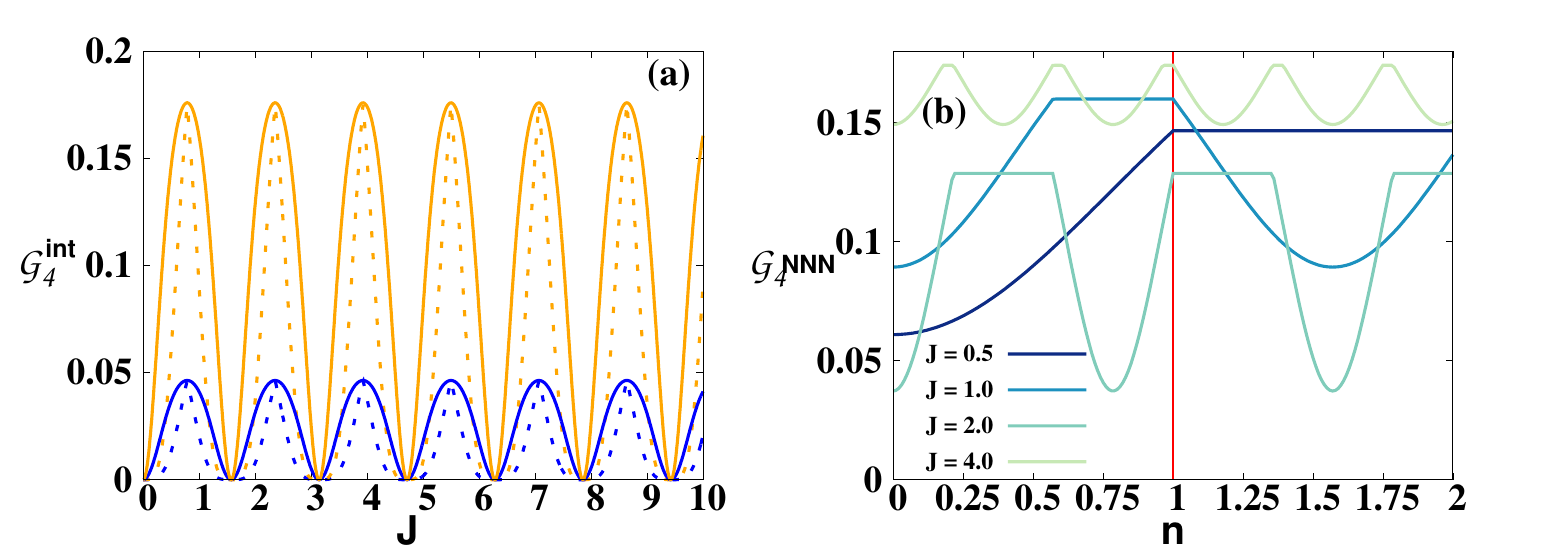}
\captionsetup{justification=Justified,singlelinecheck=false}
    \caption{Generation of genuine multimode entanglement in a four-mode circularly coupled waveguide setup. (a). Variation of GGM, $\mathcal{G}_{4}^{\text{int}}$ (ordinate) against the coupling strength $J$ (abscissa) for nearest-neighbor NN (dashed line) and next-nearest-neighbor NNN (solid lines) coupling \textcolor{black}{with $n = 1$, i.e., NNN coupling strength is the same as that of the NN coupling}.  The squeezing strength of the initial state is taken to be $s = 0.5$ (dark (blue)) and $s = 1.0$ (light (orange)). (b). $\mathcal{G}_{4}^{\text{NNN}}$ (ordinate) is plotted against the ratio of the NNN and NN coupling strengths, $n$ (abscissa). Here the increase of $J$ ($J = 0.5$, $J = 1.0$, $J = 2.0$, and $J = 4.0$) is represented from dark to light lines. The initial squeezing strength is set to $s = 1.0$. \textcolor{black}{The vertical line at $n = 1$ indicates that all the curves corresponding to a different value of $J$ attains their common maximum at that point.} All the axes are dimensionless.}
    \label{fig:4-mode}
\end{figure*}

Before going into the results concerning circular waveguides with an arbitrary number of modes, let us first investigate the situation involving a small number of modes. Such analysis can also illustrate the benefit of LR couplings for producing genuine multimode entanglement (quantified by the generalized geometric measure), with the addition of higher-order couplings one by one.
\subsection{Circular waveguide with four modes}
\label{subsec:4-mode_LR}

Let us consider a four-mode circular arrangement of waveguides, where, in addition to the NN coupling, the next-nearest-neighbor (NNN) coupling is also introduced. Before considering the situation with both NN and NNN couplings, let us first concentrate on the dynamics of multimode entanglement in the model with only NN coupling.

\subsubsection{Waveguide with nearest-neighbor couplings}
\label{subsubsec:4-mode_NN}

Let us consider the Hamiltonian for the four-mode waveguide system given in Eq. \eqref{eq:H_n} with $N = 4$ by setting $n_i = 0$ (for $i \geq 2$) which simulates only the NN couplings. By taking the initial state of the system as $\ket{\psi_s}\bigotimes\ket{0}^{\otimes3}$, whose corresponding covariance matrix is given by Eq. \eqref{eq:initial_cov-mat}, GGM is determined by finding  the symplectic eigenvalues of the  reduced covariance matrices - (i) single mode: $\Xi_f^i$ with $i  =1, \dots, 4$ and (ii) two-mode: $\Xi_f^{1j}$ with $2\leq j\leq 4$. It is observed that for each such reduced covariance matrix, there is only one symplectic eigenvalue which is not equal to $1/2$. We represent such symplectic eigenvalues of the bipartitions as $\mathbf{v} = \{\nu_i, \dots, \nu_{1j}, \dots\}$. Therefore, GGM reduces to
\begin{equation}
   \mathcal{G}_4^{\text{int}} =  \mathcal{G}_4^{\text{NN}} = 1 - \max_{\mathbf v}  \Big[\frac{2}{1 + 2 \nu_k} \Big],
     \label{eq:ggm_4_NN}
 \end{equation}
 where the superscript, $\text{``int''}$, represents the maximum LR coupling considered, while the subscript is for the total number of modes, and $k$ runs over the elements of the set $\mathbf{v}$.

\textit{Notice first that the above formalism holds for any number of modes and range of couplings (e.g., NN, NNN, etc.) as we will show in the succeeding section.} 

In the four-mode scenario, we find that the  GGM is not affected by the squeezing angle, $\theta$. Additionally, as the squeezing strength increases, so does GGM, and it is periodic with respect to $J$, with the period being  $\frac{\pi}{2}$ (see Fig. \ref{fig:4-mode} (a)). In this scenario, it is important to note that the $\nu_k$ values are dependent on both $J$ and $s$.

\subsubsection{Model with next-nearest-neighbor couplings}
\label{subsubsec:4-mode_NNN}

The Hamiltonian for simulating both the NN and the NNN couplings in a four-waveguide system can be obtained from Eq. \eqref{eq:H_n} by setting $N = 4$, $J_1 = J$, and $J_2 = n_2 J_1 = nJ$. The method for calculating GGM is similar to that for the nearest-neighbor case and it is dependent on $s$, $J$, and $n$. The strength of the next nearest-neighbor coupling can be greater than  ($n > 1$), equal to ($n = 1$), or less than ($n<1$)  that of the NN coupling. Let us now analyze the behavior of the genuine multimode entanglement with time and compare it with the scenario involving only NN couplings. To study it, we compute $\mathcal{G}_{4}^{\text{NNN}}$. The juxtaposition of $\mathcal{G}_{4}^{\text{NN}}$ and $\mathcal{G}_{4}^{\text{NNN}}$ reveals the following facts:
\begin{enumerate}
    \item Like with NN couplings, $\mathcal{G}_{4}^{\text{NNN}}$ increases with $s$ and is $\frac{\pi}{2}$-periodic with $J$.\
    
    \item On the other hand, with non-vanishing $n$, we find that $\mathcal{G}_{4}^{\text{NNN}} \geq \mathcal{G}_{4}^{\text{NN}}$ for a fixed value of $J$, although they coincide at the point where both of them reach their maximum as well as when they both are minimum.
    
    \item \textcolor{black}{Our analytical results reveal that all the symplectic eigenvalues of the reduced subsystems are sine and cosine functions of the Hamiltonian parameters, thereby leading to the oscillatory nature of GGM with respect to $n$ and $J$. Moreover, in terms of $n$, the next-nearest-neighbor coupling reads $J_2 = nJ$. For a fixed value of $J$, the NNN coupling parameter is a function of $n$ which furthermore enters into the evolution operator $\exp(-i \hat{H} t) = f(J,n)$. Therefore, the periodic behavior of GGM with $n$ is also due to the unitary evolution, similar to the oscillatory nature of $\mathcal{G}$ with $J$. Fig. \ref{fig:4-mode}(b) shows that GGM varies periodically with $n$ as well.}

    \item \textcolor{black}{Studying the variation of GGM (at a fixed $J, N$, and $s$) with $n$ helps to demonstrate the fact that the optimum generation of genuine multimode entanglement occurs when nearest-neighbor and long-range couplings have equal strength, i.e. $n = 1$, regardless of the coupling parameter $J$. Notably, from Fig. \ref{fig:4-mode}(b), we can observe that the curves exhibit periodic behavior with the variation of  $n$ for certain values of $J$. By using Fig. $2(b)$, we want to emphasize that regardless of the chosen value of $J$, GGM consistently reaches its maximum at $n = 1$ although different values of $J$ show different trends in GGM. As a consequence, it is concluded that all-to-all interactions with equal strength furnish the best possible production of genuine multimode entanglement. In other words, the enhancement of genuine multimode entanglement through LR over NN coupling is more pronounced when all the modes interact with each other equally.}

\end{enumerate}

\textbf{Note $\mathbf{2}$.} \textit{Five-mode circular waveguide system} - The GGM for the five-mode waveguide exhibits qualitatively similar properties to $\mathcal{G}_{4}^{\text{int}}$. By taking the same kind of initial state, i.e., by choosing $|\psi_s\rangle \otimes |0\rangle^{\otimes 4}$, which evolves according to $\hat{H}$ in Eq. \eqref{eq:H_n} with $J_1 = J_2 = J$, no periodicity in GGM with $J$ is observed for the nearest-neighbor case, while $\mathcal{G}^{\text{NNN}}_5$ exhibits a period of $\frac{2 \pi}{5}$. \textcolor{black}{It is important to mention here that for $N = 5$ , we can only consider up to next-nearest-neighbor couplings and also, we need to consider only the single-mode and two-mode reduced subsystems to estimate $\mathcal{G}_5^{\text{int}}$, similar to the case for the four-mode system. Therefore, $N = 4$ and $N = 5$ offer qualitatively similar insights. There, however, exists the symmetry $n \iff 1/n$ and $J \iff J/n$ (with the NN couplings interchanged with the NNN couplings) which would highly simplify the exact calculations for GGM, when $N = 5$.}

\subsection{Six-mode circular waveguide}
\label{subsec:6-mode_LR}
 \begin{figure}
    \centering
    \includegraphics[width=\linewidth]{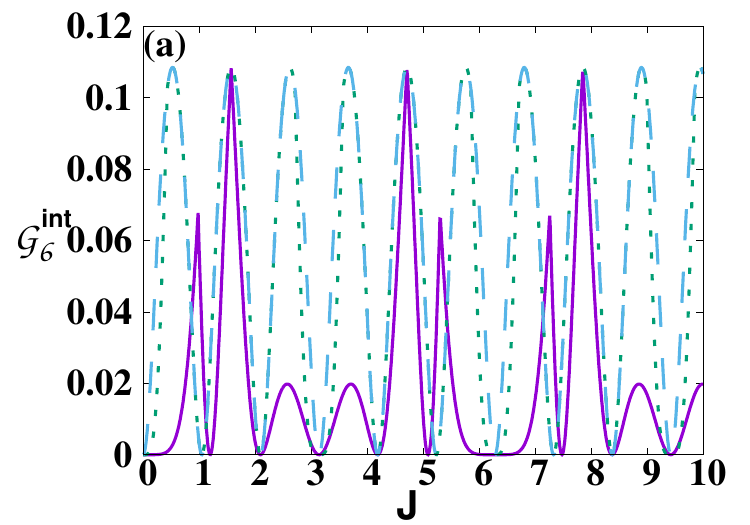}
\captionsetup{justification=Justified,singlelinecheck=false}
    \caption{Circularly coupled waveguide involving $6$-modes. The genuine six-mode entanglement $\mathcal{G}_{6}^{\text{int}}$ (ordinate)  against the coupling strength $J$ (abscissa) for $s = 1.0$ comprising NN coupling ((purple) solid line), NNN coupling ((green) dotted line) and NNNN coupling ((blue) dashed line) having equal coupling strength. All axes are dimensionless.}
    \label{fig:6-mode}
\end{figure}
We now proceed to carry out the investigation when the waveguide arrangement comprises six modes, thereby incorporating a higher level of long-range coupling like the next-to-next-nearest-neighbor (NNNN) coupling. It is interesting to find out whether LR couplings are indeed responsible for creating genuine multimode entanglement even in the presence of weak coupling strengths. The Hamiltonian for the evolution, in this case, can be realized according to Eq. \eqref{eq:H_n} with $N = 6$ and $J_1 = J$, $J_2 = J_1 = J$ and $J_3 = n_3 J_1 = n J$, where the same strength of coupling for NN and NNN couplings is considered based on the observations for the four-mode waveguides. Interestingly, the maximum GGM in the NNNN coupling case is again obtained when the coupling strength is equal to that of the short-range couplings, NN and NNN. Again GGM oscillates with $J$, irrespective of short- and long-range couplings, although, unlike the four-mode scenario, the pattern of GGM changes with the introduction of LR couplings. In particular, GGM vanishes with NNNN and NNN couplings when $J$ is a multiple of $\frac{2 \pi}{6}$, which is not the case for NN couplings and the maximal value of GGM is obtained more frequently with respect to $J$ in the presence of both LR couplings as compared to that of the NN coupling. Akin to the four- and five-mode waveguide scenarios, $\mathcal{G}_{6}^{\text{NNNN}} \geq \mathcal{G}_{6}^{\text{NNN}} \geq \mathcal{G}_{6}^{\text{NN}}$ (see Fig. \ref{fig:6-mode} (a)), although the maximum value of GGM cannot be increased by the LR couplings for a fixed $J$ value. 

We observe that LR couplings with low coupling strength can indeed create more GGM than in the NN case, as is illustrated in Fig. \ref{fig:6-mode} (a). for $0 \leq J \leq 1$.

\textit{The results of circular waveguide setups with four-, five-, and six-modes strongly indicate that incorporating long-range couplings is beneficial and that the same coupling strength for all kinds of coupling provides the best genuine multimode entanglement.}

\textcolor{black}{The constructive impact of long-range interactions is also evident from Figs. \ref{fig:4-mode}(a) and \ref{fig:6-mode}(a), from which it is clear that the area under GGM curve corresponding to LR couplings is higher than that in the presence of only NN coupling. This indicates that long-range interactions can help to create higher content of genuine multimode entanglement on average, over a given period of time, compared to the Hamiltonian with NN coupling.}

\subsection{GME produced with $N$-mode circular waveguide}
\label{subsec:N-mode_LR}

Motivated by the results obtained in the previous subsections, we compute the production of genuine multimode entanglement in arbitrary modes, say,  $N$-modes arranged in a circle. The Hamiltonian for the same pertains to an $N$-mode circularly coupled waveguide system, with its input state being specified by $\ket{\psi}_{in}^{1 2 \cdots N}=\ket{\psi_s}\otimes\ket{0}^{\otimes N-1}$. Since the studies in the previous subsections display the preferable role of equal short- and long-range coupling strengths, we take all the modes to be interacting equally with each other. 

\subsubsection{Block entropy of entanglement}
\label{subsubsec:block_entropy}
\begin{figure}
    \centering
    \includegraphics[width=\linewidth]{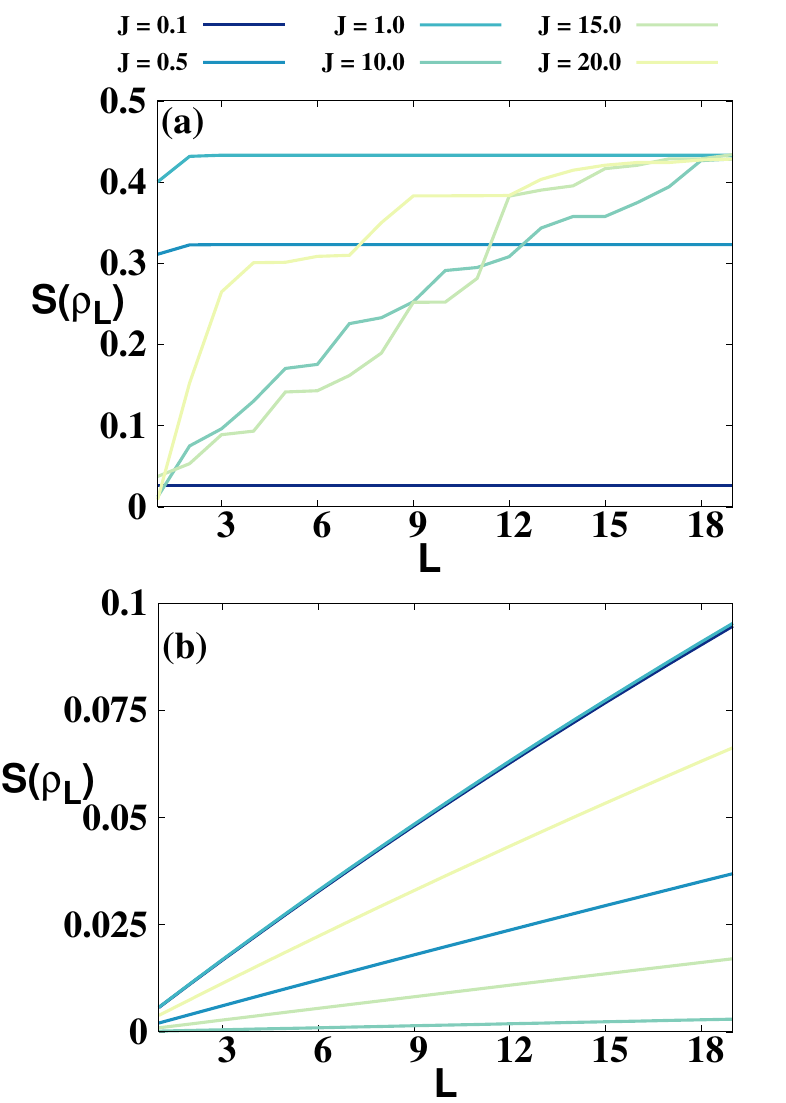}
\captionsetup{justification=Justified,singlelinecheck=false}
    \caption{Block entanglement entropy,   $S(\rho_L)$ vs the reduced system size, $L$, when the initial squeezing strength is fixed to $s = 1.0$. (a). The coupling is considered to be nearest-neighbor (NN) while in (b), couplings among all the modes are long range. Here 
 $N =40$, i.e.,   $40$ circularly  waveguide modes are coupled. From dark to light, the lines represent $J = 0.1$, $J = 0.5$, $J = 1.0$, $J = 10.0$, $J = 15.0$, and $J = 20.0$ respectively. All axes are dimensionless.}
    \label{fig:block_ent}
\end{figure}

\textcolor{black}{In order to gain some insight into the creation of genuine multimode entanglement in systems comprising all-to-all LR couplings, let us first study the behavior of the Renyi-$2$ entanglement entropy of the reduced subsystems of an $N$-mode state. Instead of quantifying the multimode entanglement geometrically,  we compare the block entropy of entanglement produced through dynamics with the nearest-neighbor coupling as well as with the long-range couplings. In particular, we look into the scaling of the entropy \cite{Wehrl_RMP_1978} for the reduced density matrices of the final state, $\ket{\psi}_f^{12...N}$, with respect to the number of subsystems comprising the reduced state. Note that we need to consider $[N/2]$ number of reduced density matrices for an $N$-mode system which are $\rho_2$, $\rho_{23}$, ..., $\rho_{23...[N/2]}$ where $\rho_{23...i}=\tr_{1,i+1,...N}\ket{\psi}_f^{12...N}\bra{\psi}_f^{12...N}$. \\
For a fixed system size, we compute the Renyi-$2$ block entropy defined as  \cite{Renyi_1961, nielsen_2010} 
\begin{eqnarray}
    S(\rho_L) = -\ln [\Tr(\rho_L^2)],
\end{eqnarray}
by varying the block size $L$, where $\rho_L = \Tr_{\bar{L}} |\psi\rangle_{f}^{12...N}\langle\psi|_{f}^{12...N}$, with $\bar{L}$ being the rest of the modes which are not included in the block, $L$. In the covariance matrix formalism, it can be simplified to $S(\rho)=\frac{1}{2}\ln(2^{2L} \det \Xi_L)$ where $\det \Xi_L$ is the determinant of the covariance matrix corresponding to an $L$-mode state $\rho_{23...L}$ \cite{Adesso_OSID_2014}. 
\\
{\it Block entropy in NN model vs LR model.} In the case of NN coupling (see Fig. \ref{fig:block_ent}(a)), it can be observed that $S(\rho_{L})$ increases with $L$ for a while, and then saturates with $L$ which increases with $J$. \textcolor{black}{Since we are dealing with a one dimensional system, this indicates that the area law, i.e., constant entanglement with \(L\), is obeyed at high block sizes.}
Moreover,  when $J \leq 1$, the block entanglement entropy saturates to different values, which again increases with $J$ while the saturation value of $S(\rho_L)$ is the same for all coupling strengths with $J > 1$.  
On the other hand, for long-range couplings with all coupling strengths being equal, as depicted in Fig. \ref{fig:block_ent}(b),  $S(\rho_L)$ always increases monotonically with $L$, thereby indicating the violation of the area law in this system and its behavior does not follow any order with respect to $J$, contrary to the NN-coupling regime. \textcolor{black}{The violation of the area law introduces a non-flat structure of the block entanglement entropy at large \(L\).} \\
\textbf{Remark}.  In Fig. \ref{fig:block_ent}, we investigate the behavior of entanglement entropy $S(\rho_{L})$ for varying partition sizes $L$ up to the system size $N/2$. We restrict our analysis to $L \leq N/2$ due to the circular arrangement of waveguides, and a similar behavior emerges for $L \geq N/2$.
 Furthermore, the non-monotone in the slope with respect to $J$ arises from the interplay between the nearest-neighbor coupling and the long-range couplings. \textcolor{black}{Note that here we are not concerned with the physical significance of the system following the area law. We aim to point out the difference in dynamics between the nearest-neighbor and long-range couplings and to gauge the subsystem which primarily contributes to the GGM, which we shall do now.}   \\
 {\it Shedding light on the computation of GGM via block entropy.} We recall that $2^{2L} \det \Xi_L = 1$ for a pure Gaussian state while it is greater than unity for a Gaussian mixed state. Thus, in the case of LR couplings, the increase in $S(\rho_{L})$  indicates that the reduced subsystems involving a larger number of modes tend towards more mixed states. It has also been established that the symplectic eigenvalues of pure Gaussian states are all equal to $1/2$ while they are greater for mixed states \cite{Adesso_OSID_2014}. Since the reduced subsystems of larger length have less purity, the symplectic eigenvalues of the single-mode reduced state contribute to GGM (since we take the maximum of $\frac{2}{1 + 2 \nu}$) as shown in Sec. \ref{subsec:N-mode_LR}, thereby shedding light on the computation of GGM.\\
 It is worth noting that the interaction strengths, $J$s, exhibit an increasing trend starting from a block of size $L=1$, which corresponds to the contribution from the $2:$rest bipartition although the calculation of block entropy does not adhere to any specific order in the case of all-to-all interaction. Consequently, the subsequent eigenvalues derived from this bipartition are responsible for GGM.  
 In the case of discrete systems, such studies have been performed \cite{Kumar_PLA_2017, Demianowicz_NJP_2021}, and we provide a similar analysis for CV systems. This is significant when the number of modes is large, since, without the knowledge of the contributing bipartition, analytical calculations would involve finding the eigenvalues of a large number of reduced density matrices, namely ($\binom{N}{1} + \binom{N}{2} + \cdots + \binom{N}{[N/2]}$). Furthermore, with the increase in the dimension of the reduced systems, it becomes analytically intractable to estimate their eigenvalues. The proof that the contribution to the genuine multimode entanglement comes from the single-mode reduced state thus helps in calculating the exact GGM produced in a system containing an arbitrary number of modes. 
 }

\subsubsection{Exact analysis of GGM in $N$-mode waveguide system}
\label{subsubsec:GGM_expresion}
Let us now derive the exact expression for GGM in this situation. Since the definition of GGM involves the Schmidt coefficients in an arbitrary number of bipartitions, the computation of GGM is hard for systems involving an arbitrary number of modes, unless some symmetry present in the system is identified. Previous configurations with four-, five-, and six-modes indicate that there is a symmetry under permutation of the modes in the evolved state,  $\ket{\psi}_{f}^{1 2 \cdots N}$, due to the circular configuration. As a consequence of this symmetry, there is only an $(N-1)$ number of different bipartitions that require to be considered. For an even number of modes i.e., $N = 2m$, the contributing bipartitions, pertaining to a given number of submodes, can be divided into two sets - one set which involves the mode in which the squeezed input state is taken, and another set that does not include the mode with the squeezed input state. Without loss of generality, if we start with a state in which the input state is plugged in the first mode, the bipartitions among the modes under study are $1:\text{rest}$, $2:\text{rest}$, $12:\text{rest}$, $23:\text{rest}$, $\cdots$, $12\dots[N/2]-1:\text{rest}$, $23\dots[N/2]:\text{rest}$ and $12\dots[N/2]:\text{rest}$, while if $N = 2m + 1$, we must consider $23\dots[N/2]+1:\text{rest}$, as an additional bipartition (here $``\text{rest}"$ in $i:\text{rest}$ denotes all the modes except $i$).  We observe that the smallest symplectic eigenvalue in $2:\text{rest}$ bipartition ultimately leads to GGM, given by
\begin{widetext}
    \begin{eqnarray}
    \nu = \frac{\sqrt{\frac{1}{4} f_1(N)-2 \sinh ^2s \left[f_2(N) \cos \left(J N\right)+\cos\left(2 J N\right)\right]+ f_3(N) \cosh 2 s}}{N^2},
    \label{eq:eignvalues}
\end{eqnarray}
\end{widetext}
where $f_1(N) = \left(N^4-4 N^2+12\right)$, $f_2(N) = \left(N^2-4\right)$, and $f_3(N) = \left(N^2-3\right)$.
The expression of GGM then takes the form as
\begin{equation}
    \label{N-mode_GGM}
    \mathcal{G}^{LR}_N=1-\frac{2}{2 \nu +1},~~~~~~~\text{for}~~N\geq4.
\end{equation}\\
 \begin{figure}
    \centering
    \includegraphics[width=\linewidth]{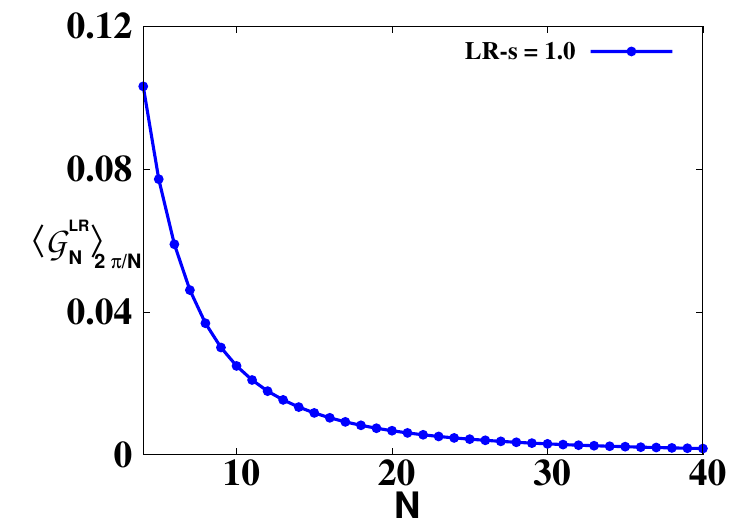}
\captionsetup{justification=Justified,singlelinecheck=false}
    \caption{$\langle{\mathcal{G}^{LR}_N}\rangle_{\frac{2 \pi}{N}}$ (ordinate)  against the number of modes $N$ (abscissa), when long-range coupling among all the modes with equal coupling strength is applied. The integration is performed a period of $\frac{2 \pi}{N}$ Here $s = 1.0$, the squeezing parameter of the initial state, $\ket{\psi_s}\otimes\ket{0}^{\otimes N-1}$. Both axes are dimensionless.}
    \label{fig:GGM_N}
\end{figure}
\textcolor{black}{It can be observed that the contributing eigenvalue $\nu$ and thus GGM is a function of $\cos \left(J N\right)$ and $\cos\left(2 J N\right)$. Therefore, a period in $J$ equalling $\frac{2 \pi}{N}$ causes the eigenvalue and hence GGM to repeat its magnitude. Based on this observation, it becomes evident that GGM possesses a period of $\frac{2 \pi}{N}$ for $N$ number of modes.}

\textcolor{black}{\textbf{Note $3$}. The problem of calculating GGM for an arbitrary number of modes is exactly solvable when we consider all-to-all interactions with equal interaction strength. In such a scenario, the Hamiltonian matrix becomes highly symmetric, $H_{p,q} = J(1 - \delta_{p,q})$ (where $\delta$ represents the Kronecker delta function), with vanishing diagonal entries and all off-diagonal entries being equal to $J$, thereby simplifying the analytical solution. To derive Eq. (\ref{eq:eignvalues}), we apply the all-to-all interaction assumption to mode number $N$ ranging from $4~\text{to}~10$ and observe a recursion relation for the eigenvalue corresponding to the $2:\text{rest}$ bipartition which contributes to GGM throughout the entire time evolution. In the computation of  GGM, the symplectic eigenvalues of the $2:\text{rest}$ bipartition only matter which can be justified through numerical simulations. To ensure the validity of the recursion relation, we verify its applicability for $N = 11, \cdots, 15$ modes and subsequently extend it to arbitrary $N$ modes.}

\textcolor{black}{\textbf{Remark}. While it is possible to solve the problem analytically for a general $n$ and $N = 3, \ldots, 6$, it becomes more challenging when $n \neq 1$ and for large N. As argued previously with $n = 1$, the Hamiltonian becomes highly symmetric, and hence the GGM can be studied analytically with certain assumption of bipartition. The Hamiltonian for $n \neq 1$ loses the particular symmetry that has been employed to calculate GGM. Moreover, for such values of $n$, the bipartition that contributes to GGM varies for different numbers of total modes, as suggested by our numerical simulations. This additional complexity makes it more difficult to obtain solutions when the eigenvalues of the relevant bipartitions cannot be obtained analytically as the system size increases. Consequently, the still symmetric nature of the Hamiltonian does not necessarily facilitate a straightforward analytical solution when considering generic values of the parameter $n$.}

\subsubsection{Effect of increasing modal number $N$ on GGM}
\label{subsubsec:GGM-N}
\textcolor{black}{We have established that an interacting Hamiltonian with long-range couplings can allow the creation of genuine multimode entanglement between an arbitrary number of modes. The magnitude of the correlations created, however, depends on the squeezing strength $s$ of the input state. Arbitrarily high squeezing cannot be created experimentally, and thus it is interesting to study the variation of GGM against the mode number $N$. For this purpose, we define the average of $\mathcal{G}_N^{\text{LR}}$ over a single period of the coupling parameter as $\langle \mathcal{G}_N^{\text{LR}} \rangle_{2\pi/N} = \frac{N}{2 \pi} \int_{0}^{2 \pi} \mathcal{G}_N^{\text{LR}} dJ$ to make it independent of $J$. This quantity also gives an insight into how much GGM can be created, on average, over a  period \(J= \frac{2\pi}{N}\).
Fig. \ref{fig:GGM_N} illustrates the variation of the average GGM against the total number of modes $N$. We observe that with an increase in $N$, the GGM created falls monotonically. 
The distribution of multimode entanglement among a larger number of modes results in a reduction in the overall content of multimode entanglement. Hence, it can be noted that this decrease in multimode entanglement is due to a finite amount of GME (multimode entanglement) that can be shared among an expanding number of modes, leading to its diminishing nature. Note, however, that for a large number of involved modes, it can be made to increase by increasing the squeezing strength $s$ of the input state although the experimental application of very high $s$ values is challenging.
}

\section{Creation of constant GGM  in waveguides with disorder}
\label{sec:disorder}

 \begin{figure*}
    \centering
    \includegraphics[width=\linewidth]{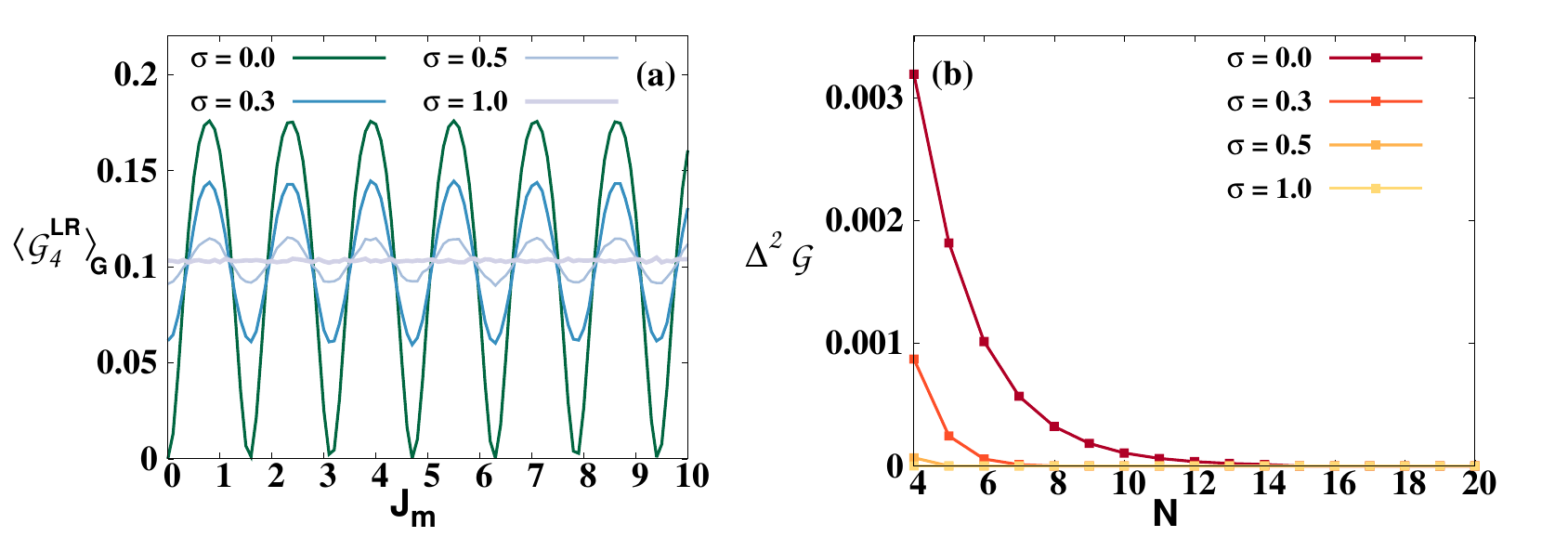}
\captionsetup{justification=Justified,singlelinecheck=false}
    \caption{Quenched averaged genuine multimode entanglement in circular waveguides coupled with disordered coupling strength. (a). The variation of the four-mode quenched GGM, $\langle \mathcal{G}^{LR}_{4} \rangle_{G}$, (ordinate) with respect to the mean coupling strength $J_m$ (ordinate).  Dark to light lines represents $\sigma = 0.0, 0.3, 0.5,$ and $1.0$ respectively.  (b). The breached GGM, $\Delta^2 \mathcal{G}$, given in Eq. (\ref{eq:fluctuating}) (ordinate) with the number of modes, $N$ (abscissa). Again, dark to light lines represents $\sigma = 0.0, 0.3, 0.5$ and $1.0$ respectively. All axes are dimensionless. }
    \label{fig:GGM_dis}
\end{figure*}

Due to the periodic nature of multimode entanglement as described in the preceding section, the method can be argued to 
have a limitation. In particular, since it collapses and revives with the variation of the coupling strength $J$,   we may end up with almost vanishing entanglement among the modes for certain values of $J$. Note that, since $J$ contains an implicit factor of time ($t$), this implies that the genuine multimode entanglement oscillates with time, thereby creating entanglement that can be used only at certain instants. A natural question at this point is how one can circumvent this feature. We indeed show that a stable (oscillation-free) multimode entangled state can be produced when the system has some imperfections. 
Given the experimental challenges in implementing couplings of a fixed strength, it is quite natural to consider that $J$ does not remain constant but oscillates around the desired value. Typically, the disorder in system parameters is responsible for the detrimental effect on the system properties, although there are certain instances in which imperfections can enhance physical characteristics \cite{Aharony_PRB_1978, Feldman_JPA_1998, Abanin_PRL_2007, Niederberger_PRL_2008, Prabhu_PRA_2011, Sadhukhan_NJP_2015, Mishra_NJP_2016, Sadhukhan_PRE_2016} like magnetization and entanglement in the modes \cite{Bera_PRB_2016, Bera_PRB_2017, Mishra_NJP_2016}. We will illustrate here other aspects of the disordered model.

To simulate such behavior, we consider a  disordered model, in which $J$ comes from a Gaussian distribution of mean $J_m$ and standard deviation $\sigma$. Here, $J_m$ is the desired coupling strength to be tuned, and a higher $\sigma$ indicates a larger oscillation around the value $J_m$, thereby measuring the strength of the disorder. We assume that the time scale taken by the disordered coupling strength to attain its equilibrium value is much larger than the implementation time, which allows us to define the \textit{quenched average GGM} over the Gaussian distribution  as 
$$\langle{\mathcal{G}^{LR}_N}\rangle_G=\frac{1}{\sqrt{2\pi}\sigma}\int_{-\infty}^{+\infty}\mathcal{G}^{LR}_N~\exp(-\frac{(J-J_m)^2}{2\sigma^2})~dJ.$$
It is observed that $\langle \mathcal{G}^{LR}_N \rangle_G$ oscillates with respect to $J_m$ with the same period $2 \pi/N$, albeit the amplitude of the oscillations decreases with increasing $\sigma$, as shown for a four-mode disordered waveguide setup in Fig. \ref{fig:GGM_dis}(a). \textcolor{black}{The presence of disorder in $J$ thus indeed leads to a smoother behavior in GGM. When the standard deviation $\sigma$ of the disorder covers at least one period of oscillation in $J$, GGM becomes more or less independent of $J_{m}$. This is because GGM in the presence of equal all-to-all coupling is periodic and has a non-zero average value. Thus its quenched average value over an entire period is constant and independent of $J_m$. It is important to note that this independence is only achieved beyond a certain threshold value of $\sigma \sim 1/N$. Thus, for large values of $N$ and small disorder, the relation $\sigma \sim 1/N$ is sufficient to completely eliminate the dependence on $J_{m}$.} 
Therefore, our findings manifest that although $\langle \mathcal{G}_{N}^{LR} \rangle_G$ decreases in comparison to the maximum GGM achieved in the ordered model, a constant GGM with lower oscillations can only be obtained when the evolution occurs according to the disordered model.

\subsection{Quantification of disorder rendered GGM stability}
\label{subsec:disorder_FOM}


The decreasing oscillations in the quenched average GGM can be quantified by the standard deviation of $\langle \mathcal{G}^{LR}_{N} \rangle_G$, which is defined as 
\begin{equation}
   \Delta^2 \mathcal{G} = \langle \langle \mathcal{G}^{LR}_{N} \rangle_{G}^{2} \rangle_{2 \pi/N} - \langle \langle{\mathcal{G}^{LR}_N}\rangle_{G}\rangle_{2 \pi/N}^{2},
   \label{eq:fluctuating}
\end{equation}
where the average is taken with respect to $J_m$ over a full cycle. We call this quantity as the \textit{breached GGM}, whose low value implies the generation of stable quenched genuine multimode entanglement. We find that this is indeed the case, i.e., the presence of disorder reduces the oscillations in the quenched average accumulated GGM. Moreover, $\Delta^2 \mathcal{G}$ decreases with the increase of $N$, as illustrated in Fig. \ref{fig:GGM_dis} (b). Our studies demonstrate that the oscillations in the quenched average GGM disappear with the increase of the disorder strength and the number of modes, although the increase of the system size has a destructive effect on the creation of GGM like the ordered system. 

\section{Conclusion}
\label{sec:conclu}

Entangled continuous variable (CV) systems are of fundamental importance in realizing a host of quantum information protocols.  Additionally, it has been demonstrated that entangled CV systems provide a key route for resolving issues with other photonic devices, such as challenges with Bell-state measurements. Therefore, designing a scheme to generate multimode-entangled states is of paramount interest. 

We demonstrated that multiple circularly coupled interacting optical waveguide modes have the potential to create highly genuine multimode-entangled states. Specifically, 
 the interacting circular waveguide can create a genuinely multimode entangled (GME) state, the entanglement being measured by using generalized geometric measure (GGM), from a squeezed or squeezed coherent state in a single mode that is product with vacuum states in the other modes. 
We point out that we have considered an experimentally feasible configuration for our study. The waveguide arrays proposed in this work can be fabricated using direct femto-second laser inscription. 
Waveguide configurations are appropriate because, unlike bulk optical elements, the propagation losses in these systems can be quite low. Additionally, the parametric down-conversion process can be used to generate the squeezed state that we have considered as the input.

We analyzed the impact of different ranges of coupling on the generation of a GME state from the product initial state. We illustrated how the incorporation of long-range couplings constructively affects the process. Specifically, long-range couplings help in generating higher genuine multimode entanglement for a fixed strength of coupling constant, compared to the circular waveguide setup with only nearest-neighbor coupling, even though the maximum value of GGM remains constant for both long-range and nearest-neighbor couplings. When the order of the long-range coupling is such that all the modes interact equally with each other, we analytically found the GGM, which varies periodically with the coupling strength.
We noticed that GGM content can be increased with an increase in the squeezing strength in the input modes. 
The benefit of LR couplings can be argued through the area under the GGM curve, which clearly shows a higher value for LR couplings, than that for waveguide modes coupled with short-range couplings.  We noted, however, that the genuine multimode entanglement generated decreases with an increase in the number of interacting modes, thereby indicating a complementary relation between system size and range of couplings.

One of the drawbacks of generating multimode entanglement via such a setup is that its magnitude oscillates with time and thus is unsuitable for utilization in protocols that require states with a certain value of entanglement. To circumvent this unwanted characteristic, we showed that the presence of disorder in the coupling between the modes of the waveguides can be useful. Starting from a product state, when the system evolves according to the circular waveguide Hamiltonian in which mode-couplings are chosen randomly from a Gaussian distribution of a fixed mean and standard deviation, with a higher standard deviation representing greater disorder in the setup, we calculated the quenched average GGM. Our results indicated that for a sufficient strength of disorder, the multimode entanglement ceases to oscillate and saturates to a fixed quenched average value. Although the quenched average GGM can never reach the maximum possible value, which can be achieved in the absence of disorder, its constant magnitude can help in its utilization in information processing tasks. In summary, our results of the disordered model used in the evolution operator are an addition to the generic physical systems, and possibly the first in photonic waveguides, which report the beneficial effect of disorder for generating genuine multimode entanglement.

Apart from the generation of genuine multimode entanglement, we showed that such a process is able to create entanglement in each bipartition. For nearest-neighbor coupling, the block entropy increases with the increase of the block length for a while and then saturates, while in the case of long-range coupling, it keeps increasing. The observation is also in good agreement with the way GGM expression is obtained in the presence of long-range coupling.

Looking at the possibility of realizing waveguide setups in laboratories, our method opens up the possibility of building quantum devices that require multimode entanglement. Although we have concentrated on photonic waveguides, our findings also apply to the coupled-cavity arrays and micro-ring resonator devices. \cite{Hartmann_Nat_2006}.

\section{Acknowledgement}
\label{sec:acknowledgement}
TA, AP, RG and ASD acknowledge the support from the Interdisciplinary Cyber-Physical Systems (ICPS) program of the Department of Science and Technology (DST), India, Grant No.: DST/ICPS/QuST/Theme- 1/2019/23. 
 AR gratefully acknowledges a research grant from Science and Engineering Research Board (SERB), Department of Science and Technology (DST), Government of India (Grant No. CRG/2019/005749) during this work. 

\appendix
\section{Primer on CV-systems}
\label{app:CV}
A continuous variable  system is characterized by quadrature variables, such as $\hat{X}$ and $\hat{P}$, which are canonically conjugate with each other \cite{Serafini_2017, Braunstein_RMP_2005}. Such observables possess an infinite spectrum and their eigenstates constitute the basis for the infinite-dimensional Hilbert space. For an $N$-mode system, the Hamiltonian comprises $2N$ parameters, $\{\hat{X}_k, \hat{P}_k\}$ (with $k = 1,2,\dots, N$), and is defined as
\begin{equation}
    \hat{H} = \frac{1}{2} \sum_{k = 1}^N (\hat{X}_k^2 + \hat{P}_k^2) = \sum_{k = 1}^N \Big(\hat{a}_k^\dagger \hat{a}_k + \frac{1}{2} \Big),
    \label{eq:CV_hamiltonian}
\end{equation}
where $\hat{a}_k^\dagger$ and $\hat{a}_k$ are the creation and annihilation operators respectively for the mode $k$ and are given in terms of the quadrature variables as
\begin{equation}
    \hat{a}_k = \frac{\hat{X}_k + i \hat{P}_k}{\sqrt{2}}, ~~~~~~~ \text{and} ~~~~~~ \hat{a}_k^\dagger = \frac{\hat{X}_k - i \hat{P}_k}{\sqrt{2}},
    \label{eq:creation-annihilation_op}
\end{equation}
with $i = \sqrt{-1}$. The creation and annihilation operators corresponding to a given mode satisfy the bosonic commutation relation, $[\hat{a}_k^\dagger, \hat{a}_k] = -1$. We can define a quadrature vector, $\hat{R} = (\hat{X}_1, \hat{P}_1, \dots, \hat{X}_N, \hat{P}_N)^T$, to rewrite the commutation relation more succinctly as
\begin{equation}
    \left[\hat{R}_k,\hat{R}_l\right]=i \mathcal{M}_{kl}\quad \text{with} ~~  \mathcal{M} = \bigoplus\limits_{j=1}^{\mathcal{N}} \Omega_j.
  \label{eq:CV_commutation}
  \end{equation}
 Here, $\mathcal{M}$ represents the $\mathcal{N}$-mode symplectic form, and $\Omega_j$, for a single mode, is given by
  \begin{equation}
      \quad \Omega_j=\begin{pmatrix}
			0 & 1\\
			-1 & 0 
		\end{pmatrix} \forall j.
  \label{eq:CV_omega}
	\end{equation}
 Out of the plethora of CV quantum states, Gaussian states constitute the most widely studied class of states \cite{ferraro2005,Weedbrook_RMP_2012}. Such states are the ground and thermal states of Hamiltonians which are at most quadratic functions of the quadrature variables. As the name suggests, Gaussian states can be completely characterized by their first and second moments, encapsulated respectively by the displacement vector $\bold{d}$ and the covariance matrix $\Xi$, in the following way:
 \begin{eqnarray}
		&& d_k=\expval{\hat{R}_k}_{\rho}, \\
  \label{eq:CV_disp}
		\Xi_{kl}&&= \frac{1}{2}\expval{\hat{R}_k\hat{R}_l+\hat{R}_l\hat{R}_k}_{\rho}-\expval{\hat{R}_k}_{\rho}\expval{\hat{R}_l}_{\rho}.
  \label{eq:CV_cov}
	\end{eqnarray}
 Here, $\rho$ denotes the $N$-mode Gaussian state under consideration and $\Xi$ is a real, symmetric, and positive definite $2N$ dimensional square matrix. Gaussian dynamics are similarly affected by second-order Hamiltonians. For analytical simplicity, we can resort to the symplectic formalism. Given any $N$-mode quadratic Hamiltonian $\hat{\mathcal{H}}$ which can be written as $\hat{\mathcal{H}}=\frac{1}{2} \hat{\xi}^\dagger H \hat{\xi}$ with $\hat{\xi}=\left(\hat{a}_1, \hat{a}_2,..., \hat{a}_N, \hat{a}_1^\dagger,..., \hat{a}_N^\dagger\right)^T$, we can construct its corresponding symplectic matrix $S_H$ as \cite{Luis_QSO_1995, Arvind_Pramana_1995, Adesso_OSID_2014}
 \begin{eqnarray}
     S_H = T^\dagger L^\dagger \exp{- i K H} L T,
     \label{eq:H_symp}
 \end{eqnarray}
 where $K, L$ and $T$ are $2N \times 2N$ matrices given by 
 \begin{eqnarray}
    && K = \begin{pmatrix}
         \mathbb{I}_N & \mathbb{O}_N \\
         \mathbb{O}_N & -\mathbb{I}_N
     \end{pmatrix}, \label{eq:K_mat} \\
    && L = \frac{1}{\sqrt{2}}\begin{pmatrix}
         \mathbb{I}_N & i \mathbb{I}_N \\
         \mathbb{I}_N & -i \mathbb{I}_N
     \end{pmatrix}, \label{eq:L_mat} \\
    && T_{jk} = \delta_{k,2j - 1} + \delta_{k + 2N, 2j} \label{eq:T_mat}.
 \end{eqnarray}
 Here, $\mathbb{I}_N$ is the $N$-dimensional identity and $\mathbb{O}_N$ is the null matrix.
 Thereafter, the evolution of the Gaussian state in terms of its displacement vector and covariance matrix is defined as \cite{Adesso_OSID_2014}
 \begin{eqnarray}
 \rho' =  e^{-i \hat{\mathcal{H}}t} \rho e^{i \hat{\mathcal{H}} t} \equiv && \bold{d}' = S_H \bold{d}, \label{eq:evolved_disp} \\
 && \Xi' = S_H \Xi S_H^T. \label{eq:evolved_cov}
 \end{eqnarray}
 
 \section{Genuine multimode entanglement for CV systems}
 \label{app:CV_GGM}
 In the discrete variable regime, a pure multipartite state, $|\psi \rangle_{1,2,\dots,N}$, is said to be genuinely entangled if it has a non-vanishing value of the generalized geometric measure (GGM) \cite{Shimony_ANYAS_1995,Barnum_JPA_2001}  defined as follows
 \begin{equation}
     \mathbb{G}(|\psi \rangle_{1,2,\dots,N}) = 1 - \max_{|\phi\rangle \in \mathcal{S}}|\langle \phi | \psi \rangle_{1,2,\dots,N}|^2,
     \label{eq:GGM_def}
 \end{equation}
 where $|\phi\rangle$ is an $N$-party pure state which is not genuinely entangled, and the Fubini Study metric is used as the distance measure \cite{Arnold_1978, Kobayashi_1996}. A simpler canonical form of GGM was derived  \cite{SenDe_PRA_2010} which reads as
 \begin{eqnarray}
      \nonumber \mathbb{G}(|\psi \rangle_{1,2,\dots,N}) = 1 - \max[\lambda_{A:B} | A \cup B && = \{1,\dots,N\}, \\
       A \cap B = \varnothing], \label{eq:GGM_simple}
 \end{eqnarray}
 where $\lambda_{A:B}$ is the maximum eigenvalue of the reduced density matrix in the $A:B$ split of the state $|\psi\rangle_{1,2,\dots,N}$. The maximization is performed over all such possible bipartitions.
 
 In the case of pure CV Gaussian systems, the genuine multimode entanglement is quantified using a similar measure \cite{Roy_PRA_2020}, defined as
 \begin{equation}
     \mathcal{G}(|\psi \rangle_{1,2,\dots,N}) = 1 - \max \mathcal{P}_m \Big[\prod_{i = 1}^m \frac{2}{1 + 2 \nu_i} \Big]_{m = 1}^{[N/2]},
     \label{eq:CV_ggm}
 \end{equation}
 where $\mathcal{P}_m$ represents all the $m$-mode reduced states corresponding to the $N$-mode pure state $|\psi \rangle_{1,2,\dots,N}$ and $\nu_i$ stand for the symplectic eigenvalues of the $m$-th reduced state. The number of such bipartitions considered is $[N/2]$ with $[x]$ denoting the integer part of $x$.

 \section{GGM for the three-mode waveguide}
 \label{sec:app_3-mode}
 The simplest Hamiltonian corresponding to Eq. \eqref{eq:H_n} is for the three-mode circular waveguide consisting of only nearest-neighbor (NN) coupling,
 \begin{equation}
     \hat{H} = \hbar J (\hat{a}_1^{\dagger} \hat{a}_2 + \hat{a}_2^{\dagger} \hat{a}_3 + \hat{a}_3^{\dagger} \hat{a}_1 + H.c.),
     \label{eq:H_3}
 \end{equation}
where we have considered the coupling strength as $J_1 = J$ and $\hbar=1$.
The symplectic eigenvalues corresponding to the evolved three-mode input state, $|\psi\rangle_{\text{in}} = |\psi_\text{s}\rangle \otimes |0\rangle^{\otimes 2}$ are given by
\begin{widetext}
\begin{eqnarray}
   && \mathbf{v}_1 = \left| \frac{1}{18} i \sqrt{16 \sinh ^2s \left(\cos 3 J + 2 \cos 6 J\right)-24 \cosh 2 s-57}\right|, \label{eq:3-mode_vA}\\
   && \text{and} \,\, \mathbf{v}_2 = \mathbf{v}_3 =  \left|\frac{1}{18} i \sqrt{8 \sinh ^2s \left(5\cos 3 J+ \cos 6 J\right)-24 \cosh 2 s-57} \label{eq:3-mode_vB}\right|,
\end{eqnarray}
\end{widetext}
where $\mathbf{v}_i$ represents the symplectic eigenvalue of the single-mode reduced states corresponding to the  $i : jk$ bipartition (for $j,k \neq i$ and $i, j, k = 1, 2, 3$). The GGM, in this case, exhibits periodic behavior with variation in $J$ at a period of $2 \pi/3$. As the initial squeezing strength of the input state increases, so does GGM. For $s = 1.0$, $\mathcal{G}_3^{\max} \approx 0.2$ at $J \approx 0.7$. In this setup, no long-range coupling is possible due to the periodic nature of the waveguide Hamiltonian.

\bibliographystyle{apsrev4-1}
\bibliography{bib}
 
\end{document}